\documentstyle[12pt,epsfig]{article}
\font\tenrm=cmr10

\font\elevenbf=cmbx10 scaled\magstep 1
\font\elevenrm=cmr10 scaled\magstep 1

\renewenvironment{thebibliography}[1]
 { \elevenrm
   \begin{list}{\arabic{enumi}.}
    {\usecounter{enumi}     \setlength{\parsep}{0pt}
     \setlength{\itemsep}{3pt} \settowidth{\labelwidth}{#1.}
     \sloppy
    }}{\end{list}}
\begin{document}
\begin{center}
\vglue 1cm
{\Large  Neutron Rich Hypernuclei in
    Chiral Soliton Model \\}
\vglue 0.4cm
{Vladimir B.~Kopeliovich$^a$\footnote{{\bf e-mail}: kopelio@inr.ru} \\
 Institute for Nuclear Research, Russian Academy of Sciences,\\ Moscow 117312, Russia}
\end{center}
{\rightskip=2pc
 \leftskip=2pc
\begin{abstract}
{\tenrm \baselineskip=10pt The binding energies of neutron rich strangeness $S=-1$ 
hypernuclei are estimated in the chiral soliton approach using 
the bound state rigid oscillator version of the $SU(3)$ quantization model. 
Additional binding of strange hypernuclei in comparison with nonstrange
neutron rich nuclei takes place at not large values of atomic (baryon) numbers,
$A=B\leq\sim 10$. This effect becomes stronger with increasing isospin of nuclides,
and for "nuclear variant" of the model with rescaled Skyrme constant $e$.
Total binding energies of $^8_\Lambda He$ and recently discovered 
$^6_\Lambda H$ satisfactorily agree with experimental data.
Hypernuclei $^7_\Lambda H$, $^9_\Lambda He$
are predicted to be bound stronger in comparison with their nonstrange analogues
$^7H$, $^9He$;  hypernuclei $^{10}_\Lambda Li$, $^{11}_\Lambda Li$,
$^{12}_\Lambda Be$, $^{13}_\Lambda Be$ etc. are bound stronger in the nuclear variant
of the model.}
\end{abstract}
 \noindent
\vglue 0.2cm
\baselineskip=14pt
\elevenrm
\section{Introduction} 
Studies of nuclear states with unusual properties --- nontrivial values of 
flavor quantum numbers (strangeness, charm or beauty), or large isospin (so called 
neutron rich nuclides)
are of permanent interest. They are closely related to the problem of existence of 
strange quark matter and its fragments, strange stars (analogues of neutron stars),
and may be important for astrophysics and cosmology.
Recently new direction of such studies, the studies of neutron rich hypernuclei with
strangeness $S=-1$, got new impact due to discovery of the
hypernucleus $^6_\Lambda H$ (heavy hyperhydrogen) by FINUDA Collaboration 
\cite{agnell2} which followed its search during several years \cite{agnell1}.

Theoretical discussion of such nuclei took place during many years, beginning with 
the work by R.H.~Dalitz and R.~Levi-Setti, \cite{dalitz} - \cite{win}, in parallel
with experimental searches \cite{juric,kubota,saha,agnell1}.
It has been noted first in \cite{dalitz} that the Lambda particle may act as additional glue for
the nuclear matter, increasing the binding energy in comparison with
nucleus having zero strangeness.
Here we confirm this observation within the chiral soliton approach (CSA).
Moreover, this effect becomes stronger for the neutron rich nuclei, with increasing
excess of neutrons inside the nucleus.

The important advantage of the CSA proposed by Skyrme \cite{skyrme}, in comparison with
traditional approaches to this problem, is its generality, 
i.e. the possibility to consider 
different nuclei on equal footing, and considerable predictive power.
(Some early descriptions of this model can be found in \cite{hosc}). 
The drawback of the CSA 
is its relatively law accuracy in describing the properties of each particular nucleus.
In this respect the CSA cannot compete with traditional approaches and models
like shell model, Hartree-Fock method, etc. \cite{dalitz} --- \cite{maj}.

The quantization of the model performed first in the $SU(2)$ 
configuration space for the baryon number one states \cite{anw}, somewhat later for configurations with
axial symmetry \cite{vkax} and for multiskyrmions \cite{irw}, allowed, in particular, to describe
the properties of nucleons and $\Delta$-isobar \cite{anw} and, more recently,
some properties of light nuclei, including so called "symmetry energy" 
\cite{ksm}\footnote{Recently the neutron rich isotope 
$^{18}B$ has been found to be unstable relative to the decay
$^{18}B \to ^{17}B +n$ \cite{spyrou}, in agreement with prediction of
the CSA \cite{ksm} }, 
and some other properties of nuclei \cite{mmw}.

The $SU(3)$ quantization of the model has been performed first within the rigid 
rotator approach \cite{gua} and also
within the bound state model \cite{cakl}.
The binding energies of the ground states of light hypernuclei have been described
in \cite{vkh} within a version of the bound state chiral soliton model \cite{wk}, 
in qualitative, even semiquantitative agreement with empirical data \cite{bmz}.
 The collective motion contributions, only, 
have been taken into account in \cite{vkh} (single particles excitations 
should be added), and special subtraction 
scheme has been used to remove uncertainties in absolute values of masses 
intrinsic to the CSA\cite{mous,mewa}.
This investigation has been extended to the higher in energy (excited) states, 
with baryon number $B=2$ and $3$, some of them may be interpreted as 
antikaon-nuclei bound states \cite{kip}.
Some of the states obtained in \cite{kip} are bound stronger than predicted originally
by Akaishi and Yamazaki \cite{akayam,yamak}. These states could overlap and appear 
in experiment as a broad enhancement, in qualitative agreement with data obtained by 
FINUDA collaboration \cite{agnell3} and more recently 
by DISTO collaboration \cite{yamaz}.

To estimate the total binding energies of neutron rich hypernuclei
we are using here one of possible $SU(3)$ quantization  models,
the rigid oscillator version of the bound state model \cite{wk} 
which seems to be the simplest one.
In section 2 our approach, the CSA, is shortly described, section 3 contains the formulas summarizing 
the CSA results for strange hypernuclei and numerical results for the binding 
energies of neutron rich hypernuclei with neutron excess $N-Z = 3$ and $4$,
atomic number $A \leq 17$. Final section contains conclusions and discussion of perspectives.
\section{Features of the CSA applied to hypernuclei}
 The CSA is based on few 
principles and ingredients incorporated in the {\it truncated}
 effective chiral lagrangian \cite{skyrme,hosc,anw}:
$$L^{eff} = -{F_\pi^2\over 16}Tr l_\mu l_\mu + {1\over 32e^2}Tr [l_\mu l_\nu]^2
+{F_\pi^2m_\pi^2\over 8}Tr \big(U+U^\dagger -2\big), \eqno (1)$$
the chiral derivative $l_\mu = \partial_\mu U U^\dagger,$ $U\in SU(2)$ or $U\in SU(3)$-
unitary matrix depending on chiral fields, 
$m_\pi$ is the pion mass, $F_\pi$- the pion decay constant known experimentally, $e$ - the only 
parameter of the model in its minimal variant proposed first by Skyrme \cite{skyrme}.

The chiral and flavor symmetry breaking term in the lagrangian density depends on kaon mass and decay constant
$m_K$ and $F_K$ ($F_K/F_\pi \simeq 1.22$ from experimental data):
$$ L^{FSB} = {F_K^2m_K^2-F_\pi^2m_\pi^2\over 24}Tr \bigl(U+U^\dagger -2\bigr) \bigl(1-\sqrt 3 \lambda_8\bigr)-$$ 
$$-{F_K^2-F_\pi^2\over 48}Tr \bigl(Ul_\mu l_\mu+ l_\mu l_\mu U^\dagger\bigr) \bigl(1-\sqrt 3 \lambda_8\bigr) 
\eqno(2). $$
This term defines the mass splittings between strange and nonstrange baryons (multibaryons), modifies some
properties of skyrmions and is crucially important in our consideration.
The whole lagrangian given by $(1),(2)$ is proportional to the number of colors of
underkying QCD, $L^{eff} \sim N_c$, which is one of justifications of the model.

The mass term $\sim F_\pi^2m_\pi^2$,
 changes asymptotics of the profile $f$ and the structure
of multiskyrmions at large $B$, in comparison with the massless case. For the $SU(2)$ case
$$ U= cos f + i\, (\vec{n}\vec{\tau}) sin\,f, \eqno (3)$$
the unit vector $\vec{n}$ depends on 2 functions, $\alpha,\;\beta$.
Three profiles $\{f,\;\alpha,\;\beta\}(x,y,z)$ parametrize the 4-component
unit vector on the 3-sphere $S^3$.

The topological soliton (skyrmion) is configuration of chiral fields, 
possessing topological charge identified with the
baryon number $B$ \cite{skyrme} (for the nucleus it is atomic number $A$: $B=A$):
$$ B= {1\over 2\pi^2}\int s_f^2 s_\alpha I\left[(f,\alpha,\beta)/(x,y,z)\right] 
d^3r, \eqno (4)$$
where $I$ is the Jacobian of the coordinates transformation, $s_f=sin\,f,\;s_\alpha=sin\,\alpha$. 
So, the quantity $B$ shows how many times the unit sphere $S^3$ is covered when 
integration over 3-dimentional space $R^3$ is made.

The important feature of the CSA is that multibaryon 
states including nuclei and hypernuclei can be considered on equal footing with the $B=1$ case.
\begin{center}
\begin{tabular}{|l|l|l|l|l|l|l|l|l|l|}                   
\hline
$B$& $ \Theta_I$& $\Theta_J$& $\Theta_F^0$ &$\Theta_S$ & $\Gamma $&$\tilde\Gamma$&$\mu_S$ &$\omega_S$ \\
\hline
$1 $& $5.55$    &$5.55 $    & $ 2.05 $   & $2.636$   &$ 4.80 $     &  $14.9$ &$3.155$ &$307$\\
\hline
$3 $&$14.4 $    &$49.7$     & $6.34$     & $8.049$     &$14.0$     &  $27.0$ &$3.069$&$289$  \\
\hline
$4 $&$16.8 $    &$78.3$     & $8.27$     & $10.47$     &$18.0$     &  $31.0$ &$2.975$&$283$  \\
\hline
$5 $&$23.5 $    &$127 $     & $10.8$     & $13.71$    &$23.8$     &  $35.0$ &$3.098$&$288$  \\
\hline
$6 $&$25.4 $    &$178 $     & $13.1$     & $16.64$     &$29.0$     &  $38.0$ &$3.125$&$287$  \\
\hline
$7 $&$28.9 $    &$221 $     & $14.7$     & $18.64$    &$32.3$     &  $44.0$ &$3.009$&$283$  \\
\hline
$8 $&$33.4 $    &$298 $     & $17.4$     & $22.15$     &$38.9$     &  $47.0$ &$3.125$&$288$  \\
\hline
$9 $&$37.8 $    &$376 $     & $20.6$     & $26.25$    &$46.3$     &  $47.5$ &$3.269$&$292$  \\
\hline
$10$&$41.4 $    &$455 $     & $23.0$     & $29.35$     &$52.0$     &  $50.0$ &$3.289$&$293$  \\
\hline
$11$&$45.2 $    &$547 $     & $25.6$     & $32.74$    &$58.5$     &  $52.4$ &$3.340$&$295$  \\
\hline
$12$&$48.5 $    &$637 $     & $28.0$     & $35.83$     &$64.1$     &  $54.6$ &$3.348$&$295$  \\
\hline
$13$&$52.1 $    &$737 $     & $30.5$     & $39.07$     &$70.2$    &  $56.8$ &$3.372$&$296$  \\
\hline
$14$&$56.1 $    &$865 $     & $33.7$    & $43.15$      &$78.2$    &  $58.9$ &$3.460$&$299$  \\
\hline
$15$&$59.8 $    &$987 $     & $36.3 $    & $46.69$     &$85.1$    &  $60.9$ &$3.498$&$301$  \\
\hline                                                   
$16$&$63.2 $    &$1110$     & $38.9$     & $50.07$     &$91.5$    &  $62.8$ &$3.517$&$302$  \\
\hline                                                   
\end{tabular}
\end{center}
{\tenrm  \bf Table 1.}  {\tenrm Characteristics of classical skyrmion configurations
which enter the nuclei --- hypernuclei binding 
energies differences. The numbers are taken from \cite{vkwz,vk01}:
 moments of inertia $\Theta$, $\Sigma$-term $\Gamma$ and $\tilde \Gamma$  -
in units $GeV^{-1}$, $\omega_S$ - in $MeV$, $\mu_S$ is dimensionless (see next sections for
explanation). All these quantities have
the lower index $B$ which is omitted for the sake of brevity. 
Parameters of the 
model $F_\pi =186\,MeV; e=4.12$ \cite{vkwz,vk01,vkh}. }\\

Minimization of the mass functional $M_{cl}$ provides 3 profiles $\{f,\alpha,\beta\}(x,y,z)$ and
allows to calculate moments of inertia $\Theta_I,\;\Theta_F$, the $\Sigma$-term 
(we call it $\Gamma$) and some other
characteristics of chiral solitons which contain implicitly   
information about interaction between baryons. 
 In Table 1 we present numerical values of the moments of inertia and other quantities 
taken from \cite{vkwz,vk01,ksh05} where analytical expressions for them
can be found as well. 
The moment of inertia $\Theta_S$ given in Tables 1 and 2 is certain combination of $\Theta_F^0$
and sigma term $\Gamma$:
$$\Theta_S = \Theta_F^0 + {1\over 4}\left({F_K^2\over F_\pi^2} -1\right)\Gamma .\eqno (5)$$
The strangeness excitation energies $\omega_S$ given in Tables 1, 2 are somewhat overestimated,
especially for nuclear variant of the model --- this is an artefact of the CSA. 
However, this overestimation is cancelled in the nuclear binding energies differences
considered below.
\begin{center}
\begin{tabular}{|l|l|l|l|l|l|l|l|l|l|}                   
\hline
$B$& $ \Theta_I$& $\Theta_F^0$ &$\Theta_S$ & $\Gamma $&$\tilde\Gamma$&$\mu_S$ &$\omega_S$ \\
\hline
$1 $& $12.8$    & $ 4.66$   & $5.893$    &$ 10.1 $   &  $19.6$ &$6.407$ &$344$\\
\hline
$6 $&$62.6 $    & $30.7$     & $38.60$    &$64.7$    &  $50.6$ &$6.728$&$334$  \\
\hline
$7 $&$69.6 $    & $34.9$     & $43.75$    &$72.5$    &  $54.4$ &$6.500$&$330$  \\
\hline
$8 $&$79.9 $    & $41.3$     & $51.97$    &$87.4$    &  $58.2$ &$6.785$&$334$  \\
\hline
$9 $&$88.9 $    & $47.1$     & $59.43$    &$101$     &  $61.7$ &$6.927$&$337$  \\
\hline
$10$&$97.4 $    & $52.6$     & $6640$     &$113$     &  $64.9$ &$6.957$&$336$  \\
\hline
$11$&$106 $    & $58.5$     & $73.88$    &$126$      &  $67.9$ &$7.038$&$337$  \\
\hline
$12$&$114 $    & $63.8$     & $80.65$     &$138$     &  $70.8$ &$7.049$&$337$  \\
\hline
$13$&$122 $    & $69.5$     & $87.94$     &$151$    &  $73.6$ &$7.102$&$338$  \\
\hline
$14$&$132 $    & $76.3$     &$96.81$     &$168$     &  $76.3$ &$7.289$&$341$  \\
\hline
$15$&$140 $    & $82.3$     &$104.5$     &$182$    &  $78.8$ &$7.353$&$342$  \\
\hline                                                   
$16$&$148 $    & $88.1$     & $112.0$     &$196$    &  $81.2$ &$7.402$&$343$  \\
\hline                                                   
\end{tabular}
\end{center}
{\tenrm \bf Table 2.} {\tenrm Same as in Table 1 for rescaled (nuclear) variant of the model
with constant $e=3.0$  \cite{ksm,ksh05}. }\\

The characteristics given in Tables 1, 2 have the following scaling properties:
$\Theta_I,\;\Theta_J,\;\Theta_F,\;\Theta_S,\;\Gamma,\;\tilde \Gamma \sim N_c$;
 $\,\mu_S,\;\omega_S \sim N_c^0\sim 1$.
The properties of the $B=2$ toroidal skyrmion, not included in Tables 1,2, have been
considered in details previously, \cite{kip} and references therein.
The rational map approximation \cite{hms} simplifies considerably calculations of various
characteristics of multiskyrmions presented in Tables 1, 2. 
\section{Spectrum of strange hypernuclei in the rigid oscillator model}  
The observed spectrum of strange multibaryon states (hypernuclei) is obtained by means of 
the $SU(3)$ quantization procedure
and depends on the quantum numbers of multibaryons and characteristics of skyrmions presented in 
Tables 1, 2.
Within the  bound state model
\cite{cakl,wk,vkh} the antikaon field is bound by the $SU(2)$ skyrmion. The mass formula takes place
$$ M = M_{cl} + \omega_S + \omega_{\bar S} + |S| \omega_S + \Delta M_{HFS} \eqno (6)$$
where strangeness and antistrangeness excitation energies 
$$\omega_S= N_c(\mu_S-1)/8\Theta_S,\;\;\omega_{\bar S}= N_c(\mu_S+1)/8\Theta_S,\eqno (7)$$
$$\Theta_S=\Theta_F^0+{1\over 4}\left({F_K^2\over F_\pi^2}-1\right)\Gamma, \quad
\mu_S = \sqrt{1+\bar m_K^2/M_0^2}\simeq 1+{\bar m_K^2\over 2M_0^2},$$ 
$$ M_0^2=N_c^2/(16\Gamma\Theta_S)\sim N_c^0,\quad \bar m_K^2=m_K^2F_K^2/F_\pi^2.
\eqno (8)$$ 
The hyperfine splitting correction to the energy of the baryon state, 
depending on hyperfine splitting
constants $c_S$, $\bar c_S$, observed isospin $I$, "strange isospin" $I_S$,
the isospin of skyrmion without added antikaons $\vec I_r$  
and the angular momentum $J$, equals in the case
when interference between usual space and isospace 
rotations is negligible or not important \cite{wk}, see also \cite{vk01,ksh05}:
$$\Delta M_{HFS} = {J(J+1)\over 2\Theta_J} 
+\frac{c_SI_r(I_r+1)- (c_S-1)I(I+1) +(\bar c_S-c_S)
I_S(I_S+1)}{2\Theta_I} \eqno (9)$$
The hyperfine splitting constants are equal
$$c_S=1- {\Theta_I\over 2\Theta_S \mu_S}(\mu_S-1) ,\qquad 
\bar c_S =1 - {\Theta_I\over \Theta_S\mu_S^2}(\mu_S-1), \eqno (10)$$
Strange isospin equals $I_S=1/2$ for $S=\mp 1$, for negative strangeness in most cases
of interest $I_S=|S|/2$ which minimizes this correction (but generally it can be not so).
We recall that body-fixed isospin $\vec I^{bf} = \vec I_r +\vec I_S$,
 \cite{wk}, \cite{vk01,ksh05}. $\vec I_r$ is quite analogous to the so called 
"right" isospin within the rotator quantization scheme \cite{gua}.
When $I_S=0$, i.e. for nonstrange states, $I=I_r$ and this formula goes over into $SU(2)$
formula for multiskyrmions.
Correction $\Delta M_{HFS}\sim 1/N_c$ is small at large $N_c$, and also for heavy 
flavors \cite{cakl,vk01}.

The mass splitting within $SU(3)$ multiplets is important for us 
here. The unknown
for the $B > 1$ solitons Casimir energy \cite{mous,mewa} cancels in the mass splittings.
For the difference of energies of states with strangeness $S$ and with $S=0$ 
which belong to multiplets with equal values of $(p,q)$-numbers ($p=2I_r$), we obtain, 
using the above expressions for the constants $c_S$ and $\bar c_S$
(it is first subtraction):
$$\Delta E (p,q; I,S; I_r,0) = |S|\omega_S + {\mu_{S}-1 \over 4\mu_{S} \Theta_{S}}
[I(I+1)-I_r(I_r+1)]+
 {(\mu_{S}-1)( \mu_{S}-2) \over 4\mu_{S}^2 \Theta_{S}}I_S(I_S+1). \eqno (11)$$
For the difference of binding energies of the hypernucleus with strangeness $S=-1$,
isospin $I=I_r-1/2$ and nonstrange nucleus with isospin $I=I_r$ 
(the neutron excess $N-Z=2I_r$) we obtain from here (second subtraction):
$$ \Delta\epsilon = \omega_{S,1}-\omega_{S,B} -{3\over 8}{\mu_{S,1}-1\over
\mu_{S,1}^2\Theta_{S,1}} +\left(I_r+{1\over 4}\right){\mu_{S,B}-1\over 4\mu_{S,B}\Theta_{S,B}}- 
{3\over 16} {(\mu_{S,B}-1) (\mu_{S,B}-2)\over \mu_{S,B}^2\Theta_{S,B}}. \eqno(12) $$ 
At $I_r=1/2$ we obtain from $(12)$ Eq. $(9)$ of previous paper \cite{vkh}.
The term $\sim (I_r+1/4)$ in Eq. $(12)$ is responsible for the additional binding of
neutron rich hypernuclei in comparison with $S=0$ neutron rich nuclei (same values of $A$ and $Z$).
The values of the quantities which enter the above formulas are shown in Tables 1, 2,
the results of calculations are presented in Tables 3 and 4.

Experimental data on total binding energies of nonstrange neutron rich nuclides
presented in first numerical columns of Tables 3 and 4 are taken from
\cite{data}.
The experimental value of binding energy of hyperhydrogen shown in Table 3,
$\epsilon (^6_\Lambda H) = 10.8 \,MeV$ is the sum of the binding energy
of $^6_\Lambda H$ relative to $^5H + \Lambda$, measured in \cite{agnell2}, 
$\epsilon (^6_\Lambda H)= (4.0\pm 1.1)MeV$ and the binding energy of $^5H$ measured
 in \cite{korsh1}, $\epsilon (^5H)\simeq 6.78\,MeV$.
\begin{center}
\begin{tabular}{|l|l|l|l|l||l|l|}
\hline
$\;\;A\,-\;_\Lambda A$&$\epsilon_{2}^{exp}$&$_\Lambda\epsilon_{3/2}^{exp}$&$\Delta\epsilon_{2,3/2}^{th}$&$\epsilon_{3/2}^{th}$
&$\Delta\epsilon_{2,3/2}^{th,*}$&$\epsilon_{3/2}^{th,*}$\\ \hline

${^6}H-{_\Lambda^6}H$      &$5.8 $ & $10.8$ & $\;\,9.0 $&$14.8$ & $11.2$ & $16 $ \\ \hline

${^8}He-{_\Lambda^8}He$    &$31.4$ & $36.0$& $\;\,3.4 $&$34.8$ & $8.9 $ & $40 $ \\ \hline

$^{10}Li-{_\Lambda^{10}}Li$&$45.3$ & $  $ & $-4.7  $&$40.6$ & $4.8$ & $50$ \\ \hline

$^{12}Be-{_\Lambda^{12}}Be$&$68.6 $ &$  $ & $-9.3  $&$59.3$ & $2.7$ & $71$ \\ \hline

$^{14}B-{_\Lambda^{14}}B$  &$85.4$ & $  $ & $-15.0 $&$70.4$ & $-1.7$ & $84$ \\ \hline

$^{16}C-{_\Lambda^{16}}C$ &$111$ & $  $ & $-18.5 $ & $92.3$ & $-3.9$ & $107$ \\ \hline
\end{tabular}
\end{center}
{\tenrm \bf Table 3.} {\tenrm The total binding energies and binding energies differences
 $\Delta\epsilon_{2,3/2}^{th}= \epsilon_{3/2} - \epsilon_2 $ between hypernuclei 
with isospin $I=3/2$ and nonstrange isotopes with $I=2,\,N-Z=4$
 (in $MeV$) for the original variant, $e=4.12$, and for the variant with
rescaled constant, $e=3$ (numbers with the $^*$). Experimental values of binding
energy are available only for $_\Lambda^8He$ \cite{juric} and 
$_\Lambda^6H$ \cite{agnell2}.}\\

The value of the binding energy of $^8_\Lambda He$ shown in Table 3 is the sum
of the $\Lambda$ separation energy $7.16 \pm 0.70\,MeV$ measured in \cite{juric}
\footnote{These data have been indicated to the author by D.E.Lanskoy.} and
the total binding energy of the $^7He$ nucleus, $\epsilon (^7He)\simeq 28.82\,MeV$.
The values marked with $^*$, $\Delta \epsilon^{th*}$ and  $\epsilon^{th,*}$, here
and in Table 4 denote the theoretical values obtained in rescaled (nuclear) variant of
the model with Skyrme constant $e=3.0$. This variant allowed to satisfactorily 
describe mass splittings of nuclear isotopes, including neutron rich nuclides, with
the mass numbers between $\sim 10$ and $30$ \cite{ksm}.
The binding energies of the ground states of hypernuclei with moderate atomic numbers
can be described within this variant of the model better than in the original variant 
($e=4.12$) \cite{vkh} (these results will be presented in next publications).
\begin{center}
\begin{tabular}{|l|l|l|l||l|l|l|}
\hline
$\;\;A\,-\;_\Lambda A$&$\epsilon_{5/2}^{exp}$&$\Delta\epsilon_{5/2,2}^{th}$&$\epsilon_{2}^{th}$
&$\Delta\epsilon_{5/2,2}^{th,*}$&$\epsilon_{2}^{th,*}$\\ \hline

${^7}H-{_\Lambda^7}H$      & $\;8  $& $\;\,15  $&$23 $ & $16.4 $ & $24  $ \\ \hline

${^9}He-{_\Lambda^9}He$    & $30.3$&  $\;\,0.1 $&$30$ & $7.0$ & $37$ \\ \hline

$^{11}Li-{_\Lambda^{11}}Li$& $45.6$& $-5.0$ & $41$ & $5.0$ & $51$ \\ \hline

$^{13}Be-{_\Lambda^{13}}Be$& $68.1 $& $-9.0$ & $59$ & $3.0$ & $71$ \\ \hline

$^{15}B-{_\Lambda^{15}}B$ &  $88.2$ & $-16 $& $72 $ & $-2.0 $ & $86 $ \\ \hline

$^{17}C-{_\Lambda^{17}}C  $& $111$& $-17  $ & $94 $ & $-2.7$ &$108$ \\ \hline
\end{tabular}
\end{center}
{\tenrm \bf Table 4.} {\tenrm Same as in Table 3 for odd atomic numbers $A$, 
hypernuclei with $I=2$ and nonstrange isotopes with $I=5/2,\;N-Z=5$.
Experimental data are not available, still}\\

The value $8\,Mev$ for the binding energy of $^7H$ is preliminary result 
published in \cite{korsh2}. It follows from Tables 4 and 3 that
hypernuclei ${_\Lambda^7}H$, ${_\Lambda^9}He$, ${_\Lambda^{11}}Li$, ${_\Lambda^{13}}Be$
and ${_\Lambda^{15}}B$ are stable relative to the decay into $\Lambda$-hyperon
and nuclei $^6H$, $^8He$, $^{10}Li$, $^{12}Be$ and $^{14}B$, in nuclear 
variant of the model. In view of our former results \cite{ksm} just the nuclear
variant of the CSA should be considered as most reliable.

We did not include the correction to the binding energies difference depending
on the spin of the nucleus $J$ by following reasons. First, this correction is small
in any case because the moment of inertia $\Theta_J$ shown in Table 1 is large,
generally $\Theta_J \sim B^2$ and $\Theta_J > B\Theta_I$. 
Besides, in some cases of interest spins of nucleus and hypernucleus
coincides, and in any case the spins of neutron rich hypernuclei are not known
presently.
The decrease of values of $\Delta\epsilon_{5/2,2}^{th}$ with increasing atomic number
may be connected with limited applicability of the rational map approximation \cite{hms}
for describing multiskyrmions at larger baryon (atomic) numbers.
\section{Conclusions and prospects}
We have calculated the difference of total binding energies of neutron rich hypernucleus
with atomic, or baryon number $A$, strangeness $S=-1$, charge $Z$ 
(i.e. containing $Z$ protons), isospin $I=(N-Z-1)/2$,
and the zero strangeness nucleus with same atomic number $A$, $Z$ protons and 
$N=A-Z$ neutrons, which has isospin $I=(N-Z)/2$.
Within the chiral soliton approach this quantity contains the smallest uncertainty.

This calculation does not contain any free parameters to be fitted.
We performed calculations for two values of the Skyrme
 constant, $e=4.12$, and for $e=3.0$ (rescaled, or nuclear variant) 
which allowed to describe the mass splittings of nuclear isotopes with atomic
numbers up to $\sim 30$ \cite{ksm}. The total binding energies of the ground states
of hypernuclei with $A \geq\sim 7$ are described better with rescaled constant $e$ than it was
made previously in \cite{vkh} with the standard value $e=4.12$.
Both variants of the model provide close results for $^6_\Lambda H$ and $^7_\Lambda H$,
but for greater atomic numbers the difference becomes considerable.
Results of the rescaled nuclear variant seem to be more reliable for 
greater atomic numbers, $A \geq\sim 10$.
Calculations performed in present paper may be extended easily to 
hypernuclei with arbitrary excess of neutrons in nuclei.

The uncertainty of our estimates is considerably greater than that of traditional
methods \cite{dalitz} - \cite{maj}, several $MeV$, at least. 
For the nucleus $_\Lambda^6H$ our result for the total binding energy is
between $14.8\,MeV$ and $16\,Mev$ in comparison with the value $10.8\pm 1.1 \,MeV$ which
can be extracted from data \cite{agnell2}, see Table 3 and its discussion.
The results of \cite{dalitz} and \cite{shimura} are in much better agreement
with data. In case of the $^8_\Lambda He$ our results are in satisfactory agreement with
previously obtained data \cite{juric}.
However, the advantage of the CSA is
that it provides a general look at nuclei with different excess of neutrons 
and great variety of atomic numbers.
We hope that results presented here may be useful for planning of future experiments
aimed to find new neutron rich hypernuclei.

Hypernuclei with quantum number beauty are expected to be bound considerably stronger
than strange hypernuclei, at least by several MeV, in some cases by few tens of MeV
\cite{vkh}. For charmed hypernuclei the binding is not so strong, due to increase of electric charge of the nucleus
by unity. More detailed calculation of binding energies in the case of beauty and charm
will be presented elsewhere.

The author is indebted to D.E.Lanskoy for reading the manuscript and useful remarks and
suggestions. \\

\elevenbf{\large \bf References}
\vglue 0.01cm

\end{document}